%% file: main-local-vs-offloading-fog.tex
\documentclass[conference]{IEEEtran}
\IEEEoverridecommandlockouts
\usepackage{cite}
\usepackage{amsmath,amssymb,amsfonts}
\usepackage{algorithmic}
\usepackage{graphicx}
\usepackage{textcomp}
\usepackage{xcolor}
\usepackage[draft,marginclue,footnote]{fixme}
\usepackage{url}
\usepackage[printonlyused]{acronym}
\usepackage{multirow}
\usepackage{xspace}

\usepackage[letterpaper, left=0.63in, right=0.635in, bottom=1in, top=0.76in]{geometry}

\newcommand{\fmbb}[2][\empty]{\ifthenelse{\equal{#1}{\empty}}{\fxnote[inline]{[BB] #2}}{\fxnote[#1]{[BB] #2}}}
\newcommand{\fmbk}[2][\empty]{\ifthenelse{\equal{#1}{\empty}}{\fxnote[inline]{[BK] #2}}{\fxnote[#1]{[BK] #2}}}
\newcommand{\fmfi}[2][\empty]{\ifthenelse{\equal{#1}{\empty}}{\fxnote[inline]{[FI] #2}}{\fxnote[#1]{[FI] #2}}}
\newcommand{\fmhb}[2][\empty]{\ifthenelse{\equal{#1}{\empty}}{\fxnote[inline]{[HB] #2}}{\fxnote[#1]{[HB] #2}}}
\newcommand{\fmpk}[2][\empty]{\ifthenelse{\equal{#1}{\empty}}{\fxnote[inline]{[PK] #2}}{\fxnote[#1]{[PK] #2}}}

\newcommand{\refFig}[1]{Fig.~\ref{#1}}

\newcommand{\refSec}[1]{Sec.~\ref{#1}}
\newcommand{\refTab}[1]{Table~\ref{#1}}
\newcommand{\refEq}[1]{\eqref{#1}}

\newcommand{\ie}{i.e.,\xspace}
\newcommand{\eg}{e.g.,\xspace}

\def\BibTeX{{\rm B\kern-.05em{\sc i\kern-.025em b}\kern-.08em
    T\kern-.1667em\lower.7ex\hbox{E}\kern-.125emX}}
    \graphicspath{ {./IMAGES/} }
\begin{document}

\title{Energy Savings by Task Offloading to a Fog Considering Radio Front-End Characteristics \\
\thanks{Copyright (c) 2019 IEEE Personal use of this material is permitted. Permission from IEEE must be obtained for all other uses, in any current or future media, including reprinting/republishing this material for advertising or promotional purposes,  creating  new  collective  works,  for  resale  or  redistribution  to  servers  or  lists,  or  reuse  of  any copyrighted component of this work in other works.
The final version of record is available at http://dx.doi.org/10.1109/PIMRC.2019.8904231
}
\thanks{This research is funded by the Polish National Centre for Research and Development within the 5th Polish-Taiwanese Joint Research Programme, project FAUST no. PL-TWV/45/2018.}
}

\author{ \IEEEauthorblockN{Pawe{\l} Kryszkiewicz}
\IEEEauthorblockA{\textit{Poznan University of Technology} \\
Pozna\'{n}, Poland\\
pawel.kryszkiewicz@put.poznan.pl}
\and
\IEEEauthorblockN{Filip Idzikowski}
\IEEEauthorblockA{\textit{Poznan University of Technology} \\
Pozna\'{n}, Poland \\
filip.idzikowski@put.poznan.pl}
\and
\IEEEauthorblockN{Bartosz Bossy}
\IEEEauthorblockA{\textit{Poznan University of Technology} \\
Pozna\'{n}, Poland \\
bartosz.bossy@put.poznan.pl}
\and
\IEEEauthorblockN{Bartosz Kopras}
\IEEEauthorblockA{\textit{Poznan University of Technology} \\
Pozna\'{n}, Poland \\
bartosz.kopras@student.put.poznan.pl}
\and
\IEEEauthorblockN{Hanna Bogucka}
\IEEEauthorblockA{\textit{Poznan University of Technology}\\
Pozna\'{n}, Poland \\
hanna.bogucka@put.poznan.pl}
%
%\and
}

\author{\IEEEauthorblockN{Pawe{\l} Kryszkiewicz\IEEEauthorrefmark{1},
Filip Idzikowski\IEEEauthorrefmark{1},
Bartosz Bossy\IEEEauthorrefmark{1},
Bartosz Kopras\IEEEauthorrefmark{1}, and
Hanna Bogucka\IEEEauthorrefmark{1}}
\IEEEauthorblockA{\IEEEauthorrefmark{1}Faculty of Electronics and Telecommunications, Poznan University of Technology, Poland}}

\maketitle

\begin{abstract}
Fog computing can be used to offload computationally intensive tasks from battery powered \ac{IoT} devices.
Although it reduces energy required for computations in an \ac{IoT} device, it uses energy for communications with the fog.
This paper analyzes when usage of fog computing is more energy efficient than local computing.
Detailed energy consumption models are built in both scenarios with the focus set on the relation between energy consumption and distortion introduced by a \ac{PA}.
Numerical results show that task offloading to a fog is the most energy efficient for short, wideband links.
\end{abstract}


\section{Introduction}
\label{sec:intro}
Cloud computing has become an established paradigm to provide various services to end users, \eg  computational resources for \ac{IoT} devices \cite{Abolfazli_Cloud_2014_comsur}.
%as long as they are connected to the cloud.
The cloud is typically located hundreds of kilometers away from the end device \cite{MukherjeeEtAl2018comst}.
The corresponding communications delay and jitter can be therefore significant.
Fog computing, being set of computational resources (e.g., local servers, small data centers) located close to the end user, has been introduced to address this problem \cite{ChiangZhang2016jiot}.
At the same time fog computing can reduce required local computing and therefore simplify \ac{IoT} end devices construction and cost. 
At the same time the power required for computations at \ac{IoT} devices is reduced.
Optimization of \ac{IoT} devices \ac{EE} is particularly important when these devices are battery powered. 

\begin{comment}
In recent years,  optimization of the \ac{EE} metric has been widely investigated in the context of resource allocation in wireless communications as well as in fog networks.
In \cite{Yan_FogEE_2018_jsac,Yu_FogEE_2019_iotj,Zhang_FogEE_2019_iotj}, \ac{EE} of fog networks is maximized by resource allocation in wireless links under delay constraints. 
However, these works take only the radiated power into account, that is needed for data transmission to the \ac{FN} and power needed for computations while omitting the \ac{RF} and \ac{BB} processing consumed energy. \
On the other hand, \cite{Sarkar2016, Sarkar2018, Deng2016} aim at the whole fog network optimization.
Here below, the problem of simplified model of power consumption for communications to \ac{FN}, from the edge \ac{IoT} terminal ("thing") perspective, is addressed. \fmbk{This part was submitted}
\end{comment}

%\cite{Yan_FogEE_2018_jsac} - FN only
%\cite{Yu_FogEE_2019_iotj} - end devices only
%\cite{Zhang_FogEE_2019_iotj} - both FNS and end devices
Research on green fog networks is usually focused on delay and energy efficiency of computations carried in fog instead of the cloud \cite{Deng2016,Sarkar2018}.
%
%Previous research on energy-efficient fog computing includes various contributions and points of view.
Modelling of energy consumption of the fog network as a whole is performed in \cite{Sarkar2016, Sarkar2018, Deng2016}.
However, energy consumption modeling of mobile devices is omitted.
% Yan_FogEE_2018_jsac
On the other hand, there are articles focused on energy consumption of mobile/\ac{IoT} devices in fog computing environment.
In \cite{Dinh2017_fog} offloading decisions of a single mobile device are examined.
In \cite{Yu_FogEE_2019_iotj,You2017,Zhang_FogEE_2019_iotj} optimization of offloading decisions of multiple mobile devices is performed jointly.
Energy consumption spent on local computations and transmission is considered in \cite{Yu_FogEE_2019_iotj,You2017,Dinh2017_fog}.
Additionally, energy spent on computations in \acp{FN} is considered in \cite{Zhang_FogEE_2019_iotj}.
However, models of energy consumption for transmission used in each of these papers are simple and do not cover \ac{BB} and \ac{RF} processing required to transmit data.
To the best of our knowledge, our work is the first one to cover these aspects in the context of fog networks.

Our research question is formulated as follows:
when is it beneficial to offload computations to the fog from the perspective of energy consumption of an \ac{IoT} device?
Both solutions, \ie local computations and data offloading consume various amounts of energy.
Energy required to process data locally is compared with energy required for preparation and sending data to the fog.
Accurate modelling of wireless transmission costs is essential to properly assess the viability of offloading computations to the fog tier.
First, we provide detailed models of power spent by end devices on transmission.
The radio transmission chain is divided into separate elements.
Significant attention is devoted to power consumption and nonlinear distortion introduced by a \ac{PA}.
It can consume significant amount of power and introduce significant distortion \cite{Nossek2011,Desset2012,Li2007_energy_consumption}.
Both effects are modeled considering soft limiter nonlinearity model and power consumption of class B \ac{PA}.
Second, the optimal operating point of the \ac{PA} is derived and used assuming \ac{SINR} maximization.
We take video surveillance as an application scenario.
It consists of one or multiple \ac{IoT} devices, \ie high-bitrate cameras utilizing
\ac{TDMA}.
We assume \ac{OFDM} for transmission which is a state-of-the-art technology in modern wireless communications, e.g, used in WiFi and \ac{LTE}.
Our results confirm that the \ac{PA} consumes significant part of link power.
This is particularly evident for long distances and narrow transmission bandwidths.
Fog computing is the most advantageous for short links with wide bandwidth available.
To the best of our knowledge, this work is the first one to consider detailed power model of an \ac{IoT} device in the context of offloading computations to fog or performing them locally.

The rest of this paper is organized as follows. 
The considered fog network architecture is presented in \refSec{sec:fog-application} along with the system model including the power consumption model of the local computations as well as in the case of offloading computations to the Fog.
\refSec{sec:results} details numerical results.
The paper is summarized in \refSec{sec:conclusions}.

\section{Fog application for video surveillance}
\label{sec:fog-application}
%\fmpk{Opisac ten scenariusz, dlaczego fog praktyczny i jak uzyty. Moze troche o architekturze takiego fog. ZADANIE: BK/FI?}
%\fmbk{Co konkretnie jest do zrobienia? Opisać przykład użycia Fog do video surveillance z literatury \cite{Kioumourtzis2017, Chen2016a}? }

\begin{figure}[htbp]
\center{\includegraphics[width=8cm]{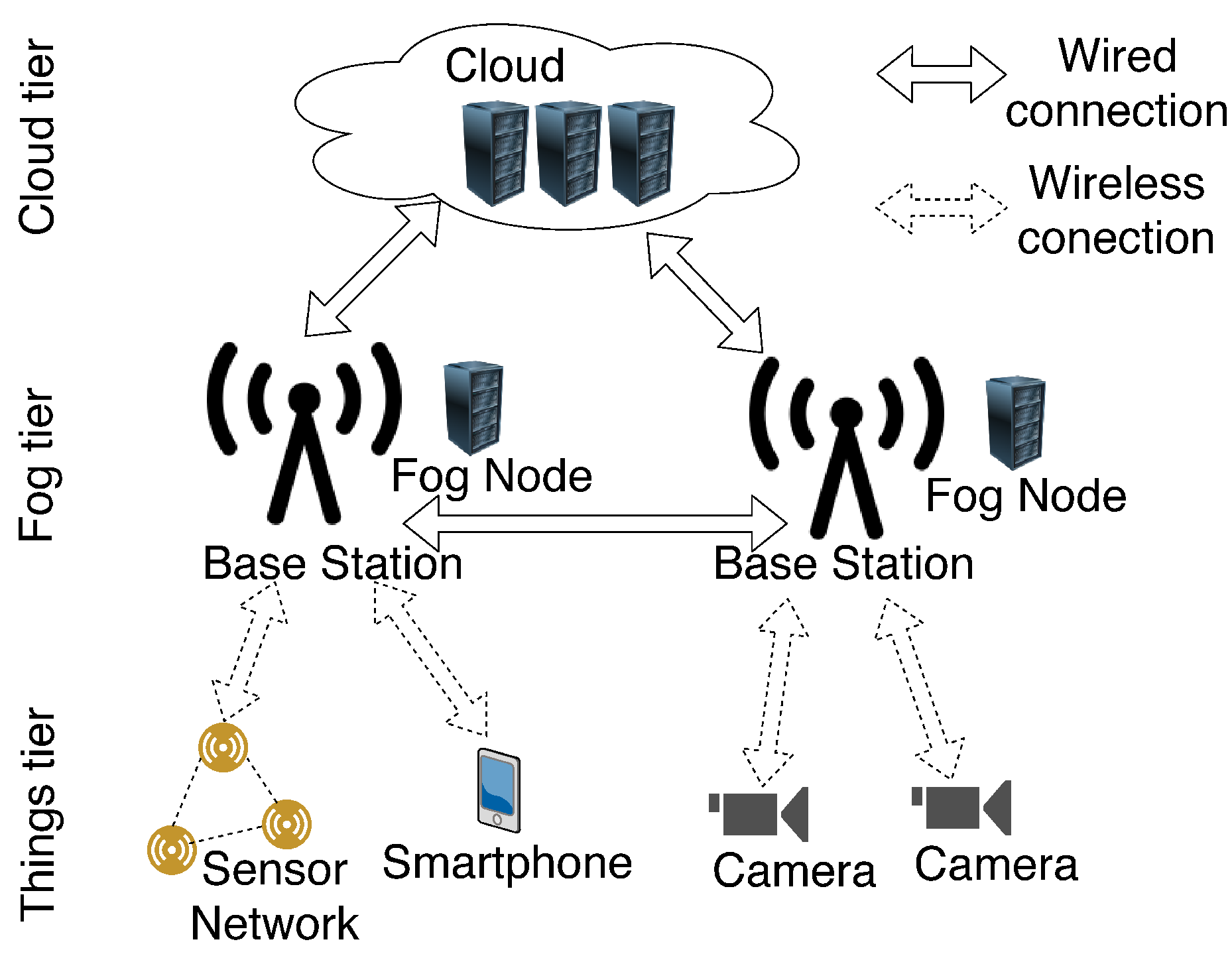}}
\caption{The considered fog network architecture.}
\label{fig:network_model}
\end{figure}

We use the fog architecture shown in \refFig{fig:network_model}.
It consists of three tiers.
The Things tier includes \ac{IoT} devices.
These devices are sources of computation requests.
The requests can be served either locally, %or in the middle tier (the Fog tier), or in the upper tier, \ie in the cloud.
or offloaded using a wireless connection to the Fog tier, \ie a nearby \ac{BS}.
%The \ac{IoT} devices use wireless transmission to communicate with a \ac{BS}.
The \ac{BS} has a short-range wired connection to a \ac{FN} which can process computational tasks sent by \ac{IoT} devices.
In addition, some computationally intensive and delay tolerant tasks can be sent to a far-away Cloud tier or distributed among many \acp{FN}.

%PK: To ponizej to powtorzenie tego co w introduction. Nic o architekturze
%The focus of this paper is set on the \ac{IoT} and wireless transmission between the things and fog tiers.
%We look at the concept of offloading computation from the \ac{IoT} device perspective -- is it more energy efficient to process the data locally or to send them to Fog?
%More precisely, we develop a detailed model for estimating the amount of energy required to transmit requests and compare it with the energy that would be spent on local computations. 
%From this point of view it does not matter whether offloaded computations are performed in a nearby \ac{FN} or a distant Cloud. 

% ================================================
% ================================================

%\section{System model}
%
%\label{sec:system-model}
In our application system, we consider $M$ video cameras, each generating constant bitrate stream that has to be processed requiring high computational load.
Face recognition is an exemplary application.
%\fmpk{MOZE INNE APLIKACJA? JAKA JEST ZLOZONOSC OBLICZENIOWA?}.

\subsection{Local computations}
In the case of local computations, the power consumption can be modeled as a function of clock frequency, voltage, and some platform specific coefficients \cite{Park2013}.
However, such a model is not directly related to the computational power measured in number of \ac{FLOPS} available.
Therefore, here, a mean, constant processing efficiency of $\Gamma=5 \textrm{ } GFLOPS/W$ is used as in \cite{Bjornson2015}.
It is inline with values reported for some smartphones, \eg Samsung Galaxy S6 \cite{Efficiency_S6_2009}.
%Knowing that the required number of floating point operations per second is $\theta \cdot R$,
The total power consumption for local calculations $P_{local}$ is defined as 
% ================================================
\begin{equation}
    P_{local}=\frac{\theta \cdot R}{\Gamma} \text{,}
    \label{eq_p_local}
\end{equation}
% ================================================   
where $R$ denotes the bitrate of video stream after compression while $\theta$ is the computational complexity coefficient determining the mean number of operations per single encoded bit required to complete the task, \eg face recognition.
Thus, $\theta \cdot R$ specifies the required computational load in \ac{FLOPS}.

\subsection{Offloading to the Fog}

The workload $\theta \cdot R$ can be offloaded to the fog network reducing energy consumption of the local device.
On the other hand, the task offloading requires the data to be transmitted via the network to the fog.
%In addition to possibly increased latency (although this is out of scope of this paper),
This forms another source of energy consumption.
The data offloading is carried in \ac{UL} of \ac{OFDM}-based system (being a common choice for modern wireless communications systems).
%of an \acs{LTE}-like network.
The downlink is neglected as a few orders of magnitude lower throughput is usually experienced. All $M$ cameras are sharing resources of a~single serving \ac{BS}, utilizing $1/M$-th of available time resources. \ac{TDMA} is the most energy efficient approach for sharing transmission media  \cite{Nossek2011}.
According to \cite{ETSI_LTE_rate}
%the \ac{LTE}
the \ac{UL} rate considering adaptive modulation and coding can be approximated by the scaled Shannon formula
% ================================================   
\begin{equation}
    R=\beta \frac{B}{M} \log_2\left(1+SINR\right) \text{,}
    \label{eq:rate_LTE}
\end{equation}
% ================================================   
where $B$ is the useful bandwidth of the transmitted signal,
%(\eg 18~MHz for 20~MHz \ac{LTE} carrier),
\acs{SINR} is \acl{SINR}, and $\beta$ is a~scaling coefficient.
In \cite{ETSI_LTE_rate} $\beta$ is equal 0.55 assuming \ac{SISO} transmission in \ac{AWGN} channel and 0.4 assuming \ac{SIMO} transmission in \ac{UL} of typical urban fast fading channel.

% ================================================   
\begin{figure}[htbp]
\centerline{\includegraphics[width=8cm]{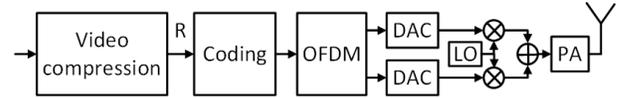}}
\caption{Diagram of the considered IoT transmitter.}
\label{fig:transmitter}
\end{figure}
% ================================================   

The diagram of the considered transmitter is shown in \refFig{fig:transmitter}.
The stream of bitrate $R$ is encoded and modulated (with power consumption detailed in \refSec{sec_coding_pow}) and transmitted using many subcarriers using \ac{OFDM} (resulting in power consumption described in \refSec{sec_OFDM_pow}).
The complex symbols at the output of \ac{OFDM} block are separated to in-phase and quadrature phase real symbols and fed to two \acp{DAC} of $N_{\mathrm{DAC}}$ bit resolution and conversion speed $f_s$, where $f_s>B$.
%, \eg in \acs{LTE} for 20 MHz carrier of $B=18$ MHz useful band the utilized sampling frequency is $f_s=15kHz\cdot 2048=30.72 MHz$.
The considered \ac{DAC} power consumption model is provided in \refSec{sec_DAC_pow}.
In order to reduce power consumption Zero-\ac{IF} architecture is considered as in \cite{Desset2012} requiring in addition only one \ac{LO} and two mixers whose power consumption models are provided in \refSec{sec_LO_pow} and \refSec{sec_mixer_pow}, respectively.
Finally, the signal at radio frequency has to be amplified in \ac{PA} and fed to the antenna.

\begin{figure}[htbp]
\centerline{\includegraphics[width=8cm]{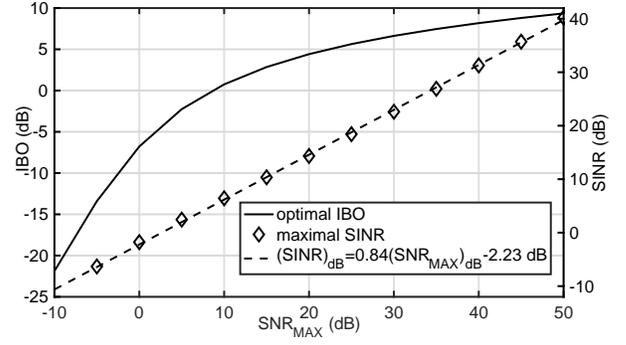}}
\caption{Optimal \ac{IBO} and \ac{SINR} value for \ac{SINR} maximization vs. $SNR_{\mathrm{MAX}}$ value.}
\label{fig:IBO_vs_SNR}
\end{figure}

In the simplest approach the higher the transmit power, the higher the received power and higher the achievable link throughput.
However, nonlinear distortion of OFDM signal passing through \ac{PA} has to be considered\cite{Nossek2011}. 
The \ac{OFDM} signal samples $x$ for sufficiently high number of subcarriers follow complex-Gaussian distribution \cite{Wei_2010_rozklad_OFDM}.
Assuming that the \ac{PA} can be modeled in \ac{BB} as a soft limiter with the input OFDM signal $x\sim \mathcal{CN} (0,\sigma^2)$, the output sample $\tilde{x}$ equals:
% ================================================
\begin{equation}
\tilde{x}=
\begin{cases}
x & \mbox{for}~~ |x|<\sqrt{P_{\mathrm{MAX}}} \\
\frac{\sqrt{P_{\mathrm{MAX}}}}{|x|}x & \mbox{for}~~ |x|\geq \sqrt{P_{\mathrm{MAX}}},
\end{cases}
\label{eq:PA_in_out}
\end{equation}
% ================================================
where $P_{\mathrm{MAX}}$ is the maximum power of a sample, \ie clipping threshold.
An \ac{IBO} defined as
% ================================================
\begin{equation}
  \mathrm{IBO}=\frac{P_{\mathrm{MAX}}}{\sigma^2}
  \label{eq_IBO_def}
\end{equation}
% ================================================
is a common measure of PA operating point, \ie how \emph{far} the mean transmission power is from the clipping power.
%s high value denotes relatively low distortion power and low energy efficiency while low values denote relatively high distortion power and high energy efficiency. 
The energy consumption of a \ac{PA} is discussed in \refSec{sec_PA_pow}. 
According to the Bussgang theorem the nonlinearly distorted signal can be decomposed into linearly down-scaled (by coefficient $\alpha$ defined in \cite{Ochiai_SINR_2000}) input signal and uncorrelated distortion $n_{\mathrm{PA}}$, \ie
% ================================================
\begin{equation}
\tilde{x}=\alpha x +n_{\mathrm{PA}}.
\end{equation}
% ================================================
This signal reaches a receiver being attenuated by the channel of a single coefficient $h$ (reflecting path loss; other effects like fading are included in coefficient $\beta$), where the noise sample $n_{\mathrm{RX}}\sim \mathcal{CN} (0,N)$ is added, resulting in
% ================================================
\begin{equation}
\tilde{y}=h\alpha x +hn_{\mathrm{PA}}+n_{\mathrm{RX}}.
\end{equation}
% ================================================
Based on derivations in \cite{Ochiai_SINR_2000,Nossek2011} it can be shown that the received \ac{SINR} value equals
% ================================================
\begin{align}
SINR&=\frac{|h|^2\alpha^2\sigma^2}{|h|^2\mathbb{E}\left[ |n_{\mathrm{PA}}|^2\right]+N}
\label{eq_SINR}
\\ \nonumber
&=\frac{\alpha^2}{1-\alpha^2-e^{-\mathrm{IBO}}+\frac{IBO}{SNR_{\mathrm{MAX}}}},
\end{align}
% ================================================
where $\alpha=1-e^{-\mathrm{IBO}}+0.5\sqrt{\pi \mathrm{IBO}}\mathrm{erfc}\left( \sqrt{\mathrm{IBO}}\right)$ and $SNR_{\mathrm{MAX}}=|h|^2P_{\mathrm{MAX}}/N$.
\ac{SINR} depends on \ac{IBO} and $SNR_{\mathrm{MAX}}$ that can be interpreted as maximum possible \ac{SINR} that can be obtained if input signal $x$ is a complex sine of power $P_{\mathrm{MAX}}$ resulting in no nonlinear distortion but maximum \ac{PA} utilization.
For \ac{SINR} maximization, the optimal \ac{IBO} value can be derived (as shown in Appendix~\ref{sec_ap_opt_IBO}) by numerically solving (\eg using the Newton method) the following equation
% ================================================
\begin{equation}
    \frac{\sqrt{\pi}}{2}\mathrm{erfc}\left(\sqrt{\mathrm{IBO}}\right)=\frac{\sqrt{\mathrm{IBO}}}{SNR_{\mathrm{MAX}}}.
    \label{eq_IBO_for_max_SINR}
\end{equation}
% ================================================
The resulting \ac{IBO} value (see \refEq{eq_SINR}) is shown in \refFig{fig:IBO_vs_SNR} for $(SNR_{\mathrm{MAX}})_{dB}\in\langle -10, 50\rangle$.
%For these \ac{IBO} values \ac{SINR} is calculated using \refEq{eq_SINR} and depicted in \refFig{fig:IBO_vs_SNR}.
Most interestingly, the maximum \ac{SINR} can be approximated in dB scale as
% ================================================
\begin{equation}
    (SINR)_{dB}\approx 0.84 (SNR_{\mathrm{MAX}})_{dB}-2.23
    \label{eq_SINR_vs_SNR_MAX}
\end{equation}
% ================================================
with maximum error lower than 0.5~dB for $(SNR_{\mathrm{MAX}})_{dB}\in\langle -10, 50\rangle$ as shown in \refFig{fig:IBO_vs_SNR}.

% ================================================
% ================================================

For a known bitrate requirement $R$, \ac{SINR} calculated from \refEq{eq:rate_LTE} can be used in \refEq{eq_SINR_vs_SNR_MAX} giving 
% ================================================
\begin{equation}
    0.84 (SNR_{\mathrm{MAX}})_{dB}-2.23=10 \log_{10}\left(2^{\frac{MR}{\beta B}} -1\right).
\end{equation}
% ================================================
After some simplifications and substitution of the  $SNR_{\mathrm{MAX}}$ definition we  obtain:
% ================================================
\begin{equation}
    P_{\mathrm{MAX}}=\frac{N}{|h|^2}10^{\frac{1}{0.84} \log_{10} \left(2^{\frac{MR}{\beta B}} -1\right)
    +\frac{2.23}{10\cdot 0.84}}.
    \label{eq_P_max_final}
\end{equation}
% ================================================
The last equation determines optimum (maximizing \ac{SINR} over \ac{IBO}) clipping power of the \ac{PA} that influences power utilization described in \refSec{sec_PA_pow}. 

The mean power consumption of a single camera for the task offloading to the fog $P_{\mathrm{off.}}$ is given as
% ================================================
  \begin{align}
    P_{\mathrm{off.}}=&P_{\mathrm{VIDEO}}+P_{\mathrm{COD}}+\frac{1}{M}P_{\mathrm{OFDM}}+\frac{2}{M}P_{\mathrm{DAC}}+
    \nonumber
    \\&+P_{\mathrm{LO}}+\frac{2}{M}P_{\mathrm{MIX}}+\frac{1}{M}P_{\mathrm{PA}},
    \label{eq:P_off}
    \end{align}  
% ================================================
where $P_{\mathrm{VIDEO}}$, $P_{\mathrm{COD}}$, $P_{\mathrm{OFDM}}$, $P_{\mathrm{DAC}}$, $P_{\mathrm{LO}}$, $P_{\mathrm{MIX}}$, and $P_{\mathrm{PA}}$ denote power consumed by video coder, redundancy coding, \ac{OFDM}, \ac{DAC}, \ac{LO}, mixer, and \ac{PA}, respectively. Division by $M$ is carried for transmitter components that are turned off when other cameras transmit their data, i.e., a component is active only $1/M$-th of time.
They are detailed in the sections below.

\subsubsection{Video compression}
\label{sec_source_coding_pow}
In order to transmit a  video signal its raw video frames have to be encoded.
H264 is one of the state-of-the-art compression standards that can be used for this purpose.
Video encoding is a computationally intensive process.
This fact together with high popularity of watching videos leads to high availability of dedicated \acp{ASIC} in the mass market \cite{Nguyen_H264_2014}.
The power consumption depends on many parameters including coder configuration, resolution, and utilized \ac{CMOS} technology.
Average power consumption is reported in \cite{Nguyen_H264_2014}, \eg $P_{\mathrm{VIDEO}}=242\textrm{ } mW$ for a Full HD stream.
The typical output bitrate is $R=6\textrm{ }Mbps$.
\subsubsection{Redundancy coding}
\label{sec_coding_pow}
The source encoded data has to be protected from transmission errors by redundancy coding, \eg turbo or \ac{LDPC}.
%, before transmission through wireless channel.
Typically, the power utilized by the coder $P_{\mathrm{COD}}$ is proportional to the bitrate $R$ \cite{Desset2012, Bjornson2015, Nossek2011}.
We assume the proportionality factor $\psi=0.1$~W/Gbps \cite{Bjornson2015} in this work , \ie $P_{\mathrm{COD}}=R\cdot \psi$.
\subsubsection{\acs{OFDM}}
\label{sec_OFDM_pow}
The computational complexity of an \ac{OFDM} transmitter is dominated by \ac{IFFT} \cite{Bogucka_2015_commMag}.
For a given transmitter bandwidth (not useful band spanned by utilized subcarriers), a subcarrier bandwidth of $\Delta f$ can be assumed giving \ac{IFFT} size $N_{OFDM}=f_s/\Delta f$.

Assuming the number of operations required for \ac{IFFT} equals $4N_{OFDM}\log_2(N_{OFDM})-6N_{OFDM}+8$ \cite{Bogucka_2015_commMag} and constant \ac{OFDM} symbol duration equal to $1/\Delta f$ (the cyclic prefix duration is neglected for simplicity) the required computational complexity equals $(4N_{OFDM}\log_2(N_{OFDM})-6N_{OFDM}+8)\Delta f$ \acp{FLOPS}.
We use $\Gamma_{MOD}$ being a constant number of operations per second per Watt to map the computational complexity to power consumption.
A value of $\Gamma_{MOD}= 120~GFLOPS/W$ is assumed for femto base stations \cite{Desset2012}.
This value can be much higher than $\Gamma$ reported for general purpose processors as the computing unit can be optimized specifically for this task.
This is related also to the utilized terminal, \ie
% ================================================
 $P_{\mathrm{OFDM}}=\left(4N_{OFDM}\log_2(N_{OFDM})-6N_{OFDM}+8\right)\Delta f \cdot \frac{1}{\Gamma_{MOD}}$.
 % ================================================
\subsubsection{\acl{DAC}}
\label{sec_DAC_pow}
The power consumption $P_{\mathrm{DAC}}$ (in Watts) of a \ac{DAC} is modeled as in \cite{Li2007_energy_consumption} as
% ================================================
\begin{equation}
    P_{\mathrm{DAC}}=V_{dd}I_{0}\left(2^{N_{\mathrm{DAC}}}-1\right)+0.5N_{\mathrm{DAC}}C_{p}f_sV_{dd}^{2},
\end{equation}
% ================================================
where $V_{dd}$ denotes power supply voltage (3~V following \cite{Li2007_energy_consumption}),
$I_{0}$ denotes unit current per least significant bit ($5\text{ }\mu A$ \cite{Li2007_energy_consumption}), $N_{\mathrm{DAC}}$ is resolution in bits of \ac{DAC} (10~bits), and $C_{p}$ is parasitic capacitance ($1\text{ }pF$ \cite{Li2007_energy_consumption}).
\subsubsection{Local Oscillator}
\label{sec_LO_pow}
Real implementations of local oscillators are reported in \cite{Rong_2016_LO, Li2007_energy_consumption}.
We assume constant power consumption of the local oscillator $P_{\mathrm{LO}}=67.5\textrm{ }[mW]$ based on \cite{Li2007_energy_consumption}.
\subsubsection{Mixer}
\label{sec_mixer_pow}
A Zero-\ac{IF} transmitter uses two mixers: one for the in-phase component and one for the quadrature component.
Based on \cite{Li2007_energy_consumption} a constant value of $P_{\mathrm{MIX}}=21\textrm{ }[mW]$ is used here for one mixer. 
\subsubsection{\acs{PA}}
\label{sec_PA_pow}
Modeling of power consumed by a \ac{PA} is a complex task depending on several factors such as \ac{PA} class, input signal distribution and specific implementation \cite{Raab_PA_2002}.
A class B \ac{PA} is considered in \cite{Nossek2011} as it provides relatively high energy efficiency (up to 78.5\%) and has relatively simple construction that is suitable for mobile devices.
Similarly, in \cite{Joung_SE_EE_PA_OFDM_2014} l-way Doherty \ac{PA} is assumed being generalization of B class \ac{PA}.
However, while \cite{Nossek2011} does not consider input signal clipping (required by the considered soft limiter \ac{PA} model), \cite{Joung_SE_EE_PA_OFDM_2014} does not consider \ac{OFDM} samples power distribution.
Therefore, the class B \ac{PA} power consumption $P_{\mathrm{PA}}$ assuming soft limiter nonlinear characteristic for complex Gaussian input signal is derived in Appendix~\ref{sec_PA_consu} giving:
% ================================================
\begin{equation}
P_{\mathrm{PA}}=\frac{2 P_{\mathrm{MAX}}}{\sqrt{\pi IBO}}\mathrm{erf}\left(\sqrt{IBO}\right),
\label{eq_power_consumed_PA_OFDM}
\end{equation}
% ================================================
where optimal \ac{IBO} is calculated solving \refEq{eq_IBO_for_max_SINR} for $SNR_{\mathrm{MAX}}$ from \refEq{eq_SINR_vs_SNR_MAX} and $P_{\mathrm{MAX}}$ is calculated based on \refEq{eq_P_max_final}.            

\section{Numerical results}
\label{sec:results}
The considered \ac{IoT} link offloading computations to the fog is analyzed using parameters summarized in \refTab{tab:parameters}.
The propagation channel amplification $10\log_{10}\left(|h|^2\right)$ assumes path loss for macro, urban base station, with 0~dBi \ac{IoT} device gain and 15~dBi base station antenna gain \cite{ETSI_LTE_rate}.
Distance between base station and \ac{IoT} device is $d$ (in km) and carrier frequency is denoted by $f$.
Noise power $N$ is modeled as thermal noise in $20^oC$ increased by 5~dB noise figure according to \cite{ETSI_LTE_rate}.
% ===========================================
\begin{table}
\centering
\caption{Parameters used in numerical evaluation.}
\label{tab:parameters}
\begin{tabular}{ll}
\hline
\textbf{Parameter} & \textbf{Value} \\ \hline
$\Gamma$ & 5 GFLOPS/W \cite{Bjornson2015}\\  %\hline
$\beta$ & 0.4 \cite{ETSI_LTE_rate} \\ %\hline
R & 6 Mbit/s \\ %\hline
$M$ & $\{1; 10\}$ \\ %\hline
$P_{VIDEO}$ & 242 mW \cite{Nguyen_H264_2014}\\ %\hline
$f_s$ & $\{15.36; 30.72\}$ MHz \cite{ETSI_LTE_rate} \\ %\hline
$B$ & $\{ 9; 18\}$ MHz \cite{ETSI_LTE_rate}\\ %\hline
$N_{OFDM}$ & $\{1024; 2048\}$ \cite{ETSI_LTE_rate}\\ %\hline
$\Gamma_{MOD}$ & 120 GFLOPS/W \cite{Desset2012}\\ %\hline
$\Delta f$ & 15 kHz \\ %\hline
$f$ & 3.5 GHz \\ %\hline
\multirow{2}{*}{$10\log_{10}\left(|h|^2\right)$} & $15\!-\!\Bigl(\! 128.1\!+\!37.6log_{10}(d\text{ }[km]) \! +$ \\
&$+\! 21log_{10}\left(\frac{f}{2\text{ }[GHz]}\right)\Bigr)$ \cite{ETSI_LTE_rate}
\\  %\hline
$(N)_{dBm}$ & $-174 +10log_{10}(B)+5$ dBm \cite{ETSI_LTE_rate} \\   %\hline
$P_{LO}$ & 67.5 mW \cite{Li2007_energy_consumption}\\ %\hline
$P_{MIX}$ & 21 mW \cite{Li2007_energy_consumption} \\ %\hline
$P_{COD}$ & $0.1\cdot R$ [mW/Mbps] \cite{Bjornson2015} \\ %\hline
$V_{DD}$ & $3\text{ }V$\cite{Li2007_energy_consumption}\\ %\hline
$I_{0}$ & $5\text{ }\mu A$\cite{Li2007_energy_consumption}\\ %\hline
$C_p$ & $1\text{ }pF$\cite{Li2007_energy_consumption}\\ %\hline
$N_{DAC}$ & 10 bit \\ \hline
\end{tabular}
\end{table}
% ===========================================

\begin{figure}[htbp]
\centerline{\includegraphics[width=8cm]{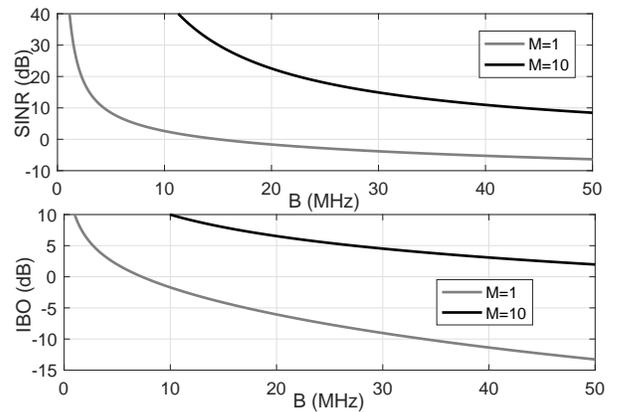}}
\caption{Optimal \ac{SINR} and \ac{IBO} for given bandwidth $B$ and constant rate $R=6\text{ }Mbps$.}
\label{fig:SINR_IBO_vs_B}
\end{figure}

The optimal \ac{SINR} changes according to \refEq{eq:rate_LTE} as shown in the top part of \refFig{fig:SINR_IBO_vs_B} under varying utilized bandwidth and fixed rate $R=6\text{ }Mbps$.
Higher bandwidth allows the same rate to be achieved by lower \ac{SINR}. Higher number of cameras sharing spectrum require higher SINR as well.
The \ac{SINR} can be used to calculate optimal parameters of the \ac{PA}, \ie $SNR_{MAX}$ using \refEq{eq_SINR_vs_SNR_MAX} and \ac{IBO} using \refEq{eq_IBO_for_max_SINR} presented in the bottom of \refFig{fig:SINR_IBO_vs_B}.
The optimal \ac{IBO} (maximizing \ac{SINR} defined by \refEq{eq_SINR}) decreases with available bandwidth achieving \ac{IBO} lower than 0~dB for B greater than 7~MHz and $M=1$.
Such low setting of \ac{IBO} results in relatively high \ac{PA} energy efficiency at the cost of relatively high nonlinear distortion power.
% ================================================
% ================================================

\begin{figure}[htbp]
\centerline{\includegraphics[width=7.6cm]{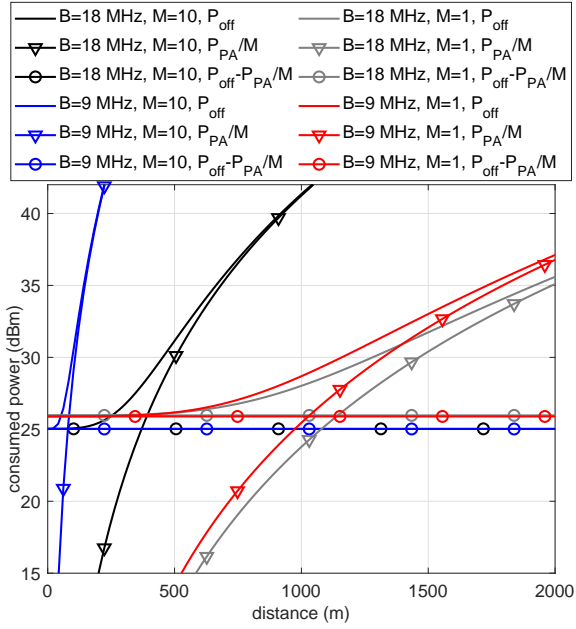}}
\caption{Power required for offloading vs. distance and available bandwidth.}
\label{fig:Pow_off_vs_distance}
\end{figure}

The required power for offloading the video stream to the fog for processing is shown in \refFig{fig:Pow_off_vs_distance} for varying distance between \ac{IoT} device and base station.
Two transmission bandwidths and 1 or 10 cameras are considered.
The power consumption for short distance communications (\eg below 20~m) equals around 26~dBm (differences below 1~dB) in each case.
This is dominated by components other than \ac{PA}.
Video compression consumes around 24~dBm.
This is inline with power consumption of WiFi cards reported in \cite{Kryszkiewicz_power_meas} where a significant constant, idle power is visible.
However, as the communications link becomes longer, the \ac{PA} power consumption becomes dominant over all power consumption components.
This phenomenon is particularly significant for narrow bandwidth and high number of cameras sharing the radio channel.
% ================================================
% ================================================

\begin{figure}[htbp]
\centerline{\includegraphics[width=7.6cm]{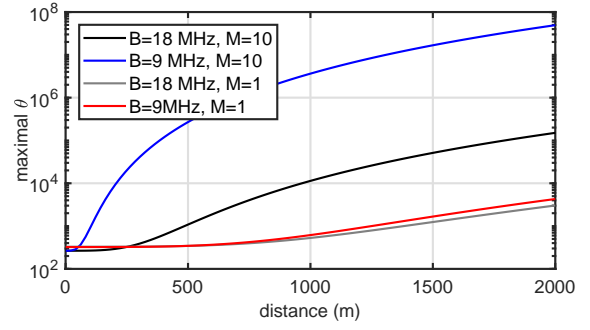}}
\caption{Maximal number of floating point operations per encoded video bit making local computations more power efficient.}
\label{fig:theta}
\end{figure}

Finally, the power consumed by an \ac{IoT} device for local computing is compared with power consumed by the same device for communications performed to offload the computation task to the fog in \refFig{fig:theta}.
The total communications power $P_{off.}$ is compared with \refEq{eq_p_local} in order to obtain the maximum number $\theta$ of \ac{FLOPS} per bit of compressed video stream that can be carried locally within the same power budget.
For short links (distance below 20~m) it is more power efficient to offload tasks to the fog if $\theta$ is greater than 320~FLOP/bit and 267~FLOP/bit for $M=1$ and $M=10$, respectively.
For longer distance the computations complexity coefficient $\theta$ has to be even higher to offload the task, \eg for $d=1000\textrm{ }m$ and one camera it equals around 620~FLOP/bit and 530~FLOP/bit for $B=$ 9~MHz and 18~MHz, respectively.
These numbers are significantly higher for 10 cameras sharing radio channel.
Maximal $\theta$ is lower for $M=10$ than for $M=1$ at distances lower than 200~m.
This is due to the fact that \ac{PA} power is low for low distances.
The constant terms from \refEq{eq:P_off} dominate $P_{off.}$ then.
Some elements (such as mixers or \acp{DAC}) can be switched off when the number of devices $M$ is bigger than 1 (see the divisions over $M$ in \refEq{eq:P_off}).
% =============================================
% =============================================

\section{Conclusions}
\label{sec:conclusions}
In the context of ever lacking energy stored in the batteries of mobile \ac{IoT} devices, we answer the question whether it is beneficial from the power consumption perspective to perform computations locally by the \ac{IoT} device or to offload (\ie send) them to the fog.
We contribute with a detailed power consumption model of \ac{IoT} device for data offloading to the fog and local computations.
Optimal operating point of the \ac{PA} is derived and used assuming \ac{SINR} maximization.
The results prove that usage of fog computing reduces \ac{IoT} device power consumption mostly for short transmission distances, wide transmission bandwidth, and relatively high local computations complexity,
\eg more than 300~FLOP per bit of compressed video stream.
Numerical values are valid for the considered, realistic system parameters.
The general trend is expected to be valid for most setups.
Verification of the analytical results
%reported in this work
against measurements of equipment on a~testbed or in a real fog network would be useful in the future.
An algorithm deciding when to offload computations to the fog is also needed.

%This requires access (or creation) of testbeds which include not only \ac{IoT} devices, but also fog nodes and ideally cloud nodes (access and management).

\appendices
%\section{\ac{SINR} at RX for softlimiter \ac{PA}}
%\label{sec_ap_SINR}
%Wyprowadzenie PK (opcja)
\section{Optimal \ac{IBO} for \ac{SINR} maximization}
\label{sec_ap_opt_IBO}
The \ac{IBO} maximizing \ac{SINR} in \refEq{eq_SINR} can be calculated by finding $z=\sqrt{IBO}$ for which \ac{SINR} derivative equals 0, \ie 
% =============================================
\begin{align}
   & \frac{\partial SINR}{\partial z}=\frac{2\alpha
    }{\left(1-\alpha^2-e^{-z^2}+\frac{z^2}{SNR_{\mathrm{MAX}}}\right)^{2}}
    \cdot \\ &
    \cdot
     \left(\frac{\partial \alpha}{\partial z}\left(\!1\!\!-\!\!e^{ -z^2}\!\!+\!\!\frac{z^2}{SNR_{\mathrm{MAX}}}\!\!\right)
        \!-\alpha ze^{-z^2}
    \!\!-\!\frac{\alpha z}{SNR_{\mathrm{MAX}}}
    \right)\!=\!0,
    \nonumber
\end{align}
% =============================================
where $\frac{\partial \alpha}{\partial z}=z e^{ -z^2}+\frac{\sqrt{\pi}}{2}\mathrm{erfc}\left( z\right)$.
Observing that the denominator is always positive, the root of the nominator has to be found.
The following equation is obtained after omission of denominator and substitution of $\frac{\partial \alpha}{\partial z}$ and $\alpha$:
% =============================================
\begin{equation}
\left(\!1\!-\!e^{ -z^2}\!-\!z^2e^{ -z^2}\right)
\left( \frac{\sqrt{\pi}}{2}\mathrm{erfc}\left( z\right)-\frac{z}{SNR_{\mathrm{MAX}}}\right) =0.
\end{equation}
% =============================================
In the above product the first factor is always positive for $z>0$.
This stems from the fact that it equals 0 for $z=0$ and its derivative, \ie $2z^3 e^{-z^2}$, is positive for any $z>0$.
Therefore, the second factor has to be equal 0 resulting in \refEq{eq_IBO_for_max_SINR}.

\section{Class B \ac{PA} power consumption for complex Gaussian input signal}
\label{sec_PA_consu}
For the input \ac{PA} signal $x$ the output signal $\tilde{x}$, described by \refEq{eq:PA_in_out}, achieves maximum power $P_{\mathrm{MAX}}$.
Consumed power $\hat{P}_{\mathrm{PA}}$ of a class B \ac{PA} transmitting sine wave of mean power $p$ is given as 
% =============================================
\begin{equation}
   \hat{P}_{\mathrm{PA}}= \frac{4}{\pi}\sqrt{pP_{\mathrm{MAX}}}
\end{equation}
% =============================================
following \cite{Raab_PA_2002, Nossek2011}.
It has to be averaged over experienced instantaneous power distribution for any other transmitted signal.
Each \ac{OFDM} symbol sample $x$ can be well approximated by a complex normal distribution with variance $\sigma^2$, \ie $x\sim \mathcal{CN} (0,\sigma^2)$.
Therefore, amplitude $z=|x|$ is Rayleigh distributed with parameter $\sigma/\sqrt{2}$ resulting in probability density function
% =============================================
\begin{equation}
    f_z(z)=\frac{2z}{\sigma^{2}} \exp{\left(-\frac{z^2}{\sigma^2}\right)}.
\end{equation}
% =============================================
The mean \ac{PA} power consumption is calculated as follows:
% =============================================
\begin{equation}
   P_{\mathrm{PA}}=E\left[ \hat{P}_{\mathrm{PA}}\right]= \frac{4}{\pi}\sqrt{P_{\mathrm{MAX}}}\int_0^{\infty}\left|\tilde{x}\right|f_z(z) dz.
\end{equation}
% =============================================
It can be rewritten utilizing \refEq{eq:PA_in_out} as
% =============================================
\begin{equation}
   P_{\mathrm{PA}}= \frac{4}{\pi}\sqrt{P_{\mathrm{MAX}}}\left(
   \int_0^{\sqrt{P_{\mathrm{MAX}}}}\!\!\!\!\!\!\!\!\!\!\!\!\!\!\!\!\!zf_z(z) dz+\int_{\sqrt{P_{\mathrm{MAX}}}}^{\infty}\!\!\!\!\!\!\!\!\!\!\!\!\sqrt{P_{\mathrm{MAX}}}f_z(z)dz
   \right). 
   \label{eq:OFDM-integral}
\end{equation}
% =============================================
It can be found that
% =============================================
\begin{equation}
   \int \frac{z^2}{B}\exp{\left(-\frac{z^2}{B}\right)}dz=\frac{\sqrt{\pi B}}{4} \mathrm{erf}\left(\frac{z}{\sqrt{B}} \right)-\frac{z}{2}\exp{\left(-\frac{z^2}{B}\right)}
\end{equation}
% =============================================
and
% =============================================
\begin{equation}
   \int \frac{z}{B}\exp{\left(-\frac{z^2}{B}\right)}dz=-\frac{1}{2}\exp{\left(-\frac{z^2}{B}\right)}.
\end{equation}
% =============================================
Substituting these results to \refEq{eq:OFDM-integral} and after some simplifications we obtain
% =============================================
\begin{equation}
   P_{\mathrm{PA}}= 2\sqrt{\frac{P_{\mathrm{MAX}}\sigma^2}{\pi}}\mathrm{erf}\left(\sqrt{\frac{P_{\mathrm{MAX}}}{\sigma^2}} \right)
\end{equation}
% =============================================
where $\sigma^2$ can be substituted using \refEq{eq_IBO_def} resulting in \refEq{eq_power_consumed_PA_OFDM}.

\input{acronyms.tex}
% Balance out the columns on the last page
\bibliographystyle{IEEEtran}
\bibliography{local-vs-offloading-fog}

\end{document}

%% file: acronyms.tex
% ========================================
%      ACRONYMS
% ========================================
\acrodef{ASIC}{Application-Specific Integrated Circuit}
\acrodef{AWGN}{Additive White Gaussian Noise}
\acrodef{BB}{BaseBand}
\acrodef{BS}{Base Station}
\acrodef{CMOS}{Complementary Metal-Oxide-Semiconductor}
\acrodef{DAC}{Digital-Analog Converter}
\acrodef{DVB-T}{Digital Video Broadcasting -- Terrestrial}
\acrodef{EE}{Energy Efficiency}
\acrodef{FLOPS}{Floating point Operations Per Second}
\acrodef{FN}{Fog Node}
\acrodef{IBO}{Input Back-Off}
\acrodef{ICT}{Information and Communication Technology}
\acrodef{IF}{Intermediate Frequency}
\acrodef{IFFT}{Inverse Fast Fourier Transform}
\acrodef{IoT}{Internet of Things}
\acrodef{LDPC}{Low-Density Parity Check}
\acrodef{LO}{Local Oscillator}
\acrodef{LTE}{Long Term Evolution}
\acrodef{OFDM}{Orthogonal Frequency-Division Multiplexing}
\acrodef{PA}{Power Amplifier}
\acrodef{RF}{Radio Frequency}
\acrodef{SIMO}{Single-Input Multiple-Output}
\acrodef{SINR}{Signal-to-Interference and Noise Ratio}
\acrodef{SISO}{Single-Input Single-Output}
\acrodef{UE}{User Equipment}
\acrodef{UL}{Up Link}
\acrodef{TDMA}{Time-Division Multiple Access}